\documentstyle[prl,twocolumn,aps,graphics]{revtex}
\input{epsf}
\begin{document}
\draft
\twocolumn[\hsize\textwidth\columnwidth\hsize\csname @twocolumnfalse\endcsname
\title{Current-current correlations in a model $CuO_3$ system
 }

\author{Bhargavi Srinivasan and Marie-Bernadette Lepetit}

\address {Laboratoire de Physique Quantique, UMR 5626 du CNRS, 
Universit\'e Paul Sabatier, F-31062 Toulouse Cedex 4, France}

\date{\today}

\maketitle

\begin{abstract}
We study a 3-band extended Hubbard model on a $CuO_3$ system
 by the density matrix renormalization group (DMRG) method.
Our system has geometry
similar to that of a section of the $CuO_2$ plane of the copper-oxide
superconductors.
We have studied the  effects of nearest-neighbour
repulsions on  the current-current corre\-lations of 
two model $CuO_3$ systems.
We have calculated several types of staggered cur\-rent correlations
as a function of various 
model parameters such as interaction strengths and filling. 
We show that repulsive interactions
are clearly capable of increasing the magnitude of the
current-current correlations in both models studied. 
The long-distance behavior of the current-current correlations
is qualitatively influenced by $V$.
We show that repulsive interactions are capable of 
enhancing correlations involving at least one copper-oxygen
rung over distances comparable to the length of the
system. In some cases, $V$ decreases the exponent 
corresponding to the power-law decay, leading to a 
considerable slowing down of the long-distance decay 
of the current-current correlations, compared to that
of the Hubbard model.
\end{abstract}
\vskip1pc]

\narrowtext

\newpage
\section{Introduction}
The unusual normal-state phenomenology of the 
high $T_c$ superconductors 
constitutes an important theoretical challenge since
there is not yet a consensus in the community 
about a  consistent
microscopic theory that would reproduce all aspects of 
the phase diagram
of the cuprates. 
Angle-resolved photoemission  spectroscopy (ARPES) experiments
have revealed several properties of the so-called
pseudogap phase in the underdoped part of the
phase diagram \cite{timusk}
and it is likely that this phase holds the key 
to a general understanding of the phase diagram.
Among the scenarios proposed to explain the pseudogap
phase, we mention the 'circulating-current' order-parameter,
which breaks time-reversal symmetry and four-fold rotational
symmetry about the copper atoms, while preserving the translational
symmetry of the lattice\cite{varma}. This theory  also gives rise
to non-Fermi liquid behaviour. The 'd-density wave' (DDW) 
order-parameter
based on orbital currents \cite{ddw} has been proposed as
a hidden order parameter for the phase diagram of the 
cuprates. In Fig. 1 we present a  schematic picture of these 
circulating currents in the $CuO_2$ plane 
of the copper-oxides.

Phases similar to those with DDW or circulating-currents order-parameters
have been discussed 
previously in the literature under 
different guises. The so-called orbital antiferromagnet 
is a phase with broken  translational and time-reversal symmetries, 
described for the excitonic insulator \cite{halperin} 
and in the two-dimensional (2D)
Hubbard model \cite{schulz}. 'Flux phases' were studied in
the context of the large n limit of the Heisenberg model \cite{affleck}.
 Staggered currents were found to 
be stabilized by attractive nearest-neighbour interactions 
in the  2D Hubbard model \cite{stanescu}. 
On the other hand, $t-J$ clusters \cite{lee,leung} were
shown variationally and numerically to have a staggered flux
order-parameter. 
Recently, the doped two-leg $t-J$ ladder, studied by the
density matrix renormalization group (DMRG) method \cite{scalapino}
was found to have exponentially decaying 
rung-rung current-current correlations.
However, the staggered flux phase was found to be stabilized in the
half-filled 2-leg ladder by abelian bosonization methods\cite{marston1}. 
Further, inclusion of longer-range interactions in the
2-leg ladder (extended $t-U-V-J$ model) appears to 
 stabilize the
staggered flux phase, as seen from  DMRG calculations\cite{marston2}.

\begin{figure}
\begin{center}
\resizebox{6.0cm}{6cm}{\includegraphics{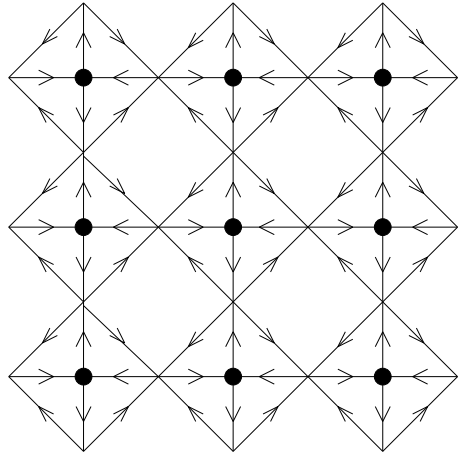}}
\end{center}
\caption{ 
Schematic picture of the circulating-current phase in 
the copper-oxides. The filled circles represent copper atoms
and the vertices oxygen atoms of the $CuO_2$ plane. The
arrows on each bond  correspond to the direction of the current
along the bond.
  }
\end{figure}

With this collection of stimulating results on flux phases in
the Hubbard and $t-J$ models, it would be interesting to study a 
system whose geometry more closely resembles that of the copper-oxides.
This is especially relevant in the light of the debate
regarding the suitability of the $t-J$
model to the copper-oxide system and in particular
the importance of an explicit treatment of the oxygen atoms
\cite{dagotto}.
There is now general agreement in the community that 
the $CuO_2$ planes play a crucial role in the high $T_c$
superconductors\cite{dagotto} and that the geometry 
of the copper-oxides is relevant to superconductivity. 
Further, several models
with effective interactions have been proposed to describe the
$CuO_2$ plane\cite{varma3,emery3,dagotto}. These studies
have emphasized the role of strong electronic correlations. Mean-field
studies\cite{varma} have indicated that it is necessary to go
beyond on-site interactions and an explicit treatment of the
nearest-neighbour interactions is necessary to stabilize the 
circulating-current phase.  An accurate treatment of the 
strongly correlated  full 2D $CuO_2$ system, though desirable,
is not presently forthcoming. Therefore, we have opted for a 
$CuO_3$ system that we believe to incorporate the crucial  geometric
ingredients of the system, including an explicit treatment 
of oxygen sites, which  can then be studied by the 
highly accurate DMRG method\cite{white}.

\begin{figure}
\begin{center}
\epsfxsize=0.8\columnwidth
\epsfbox{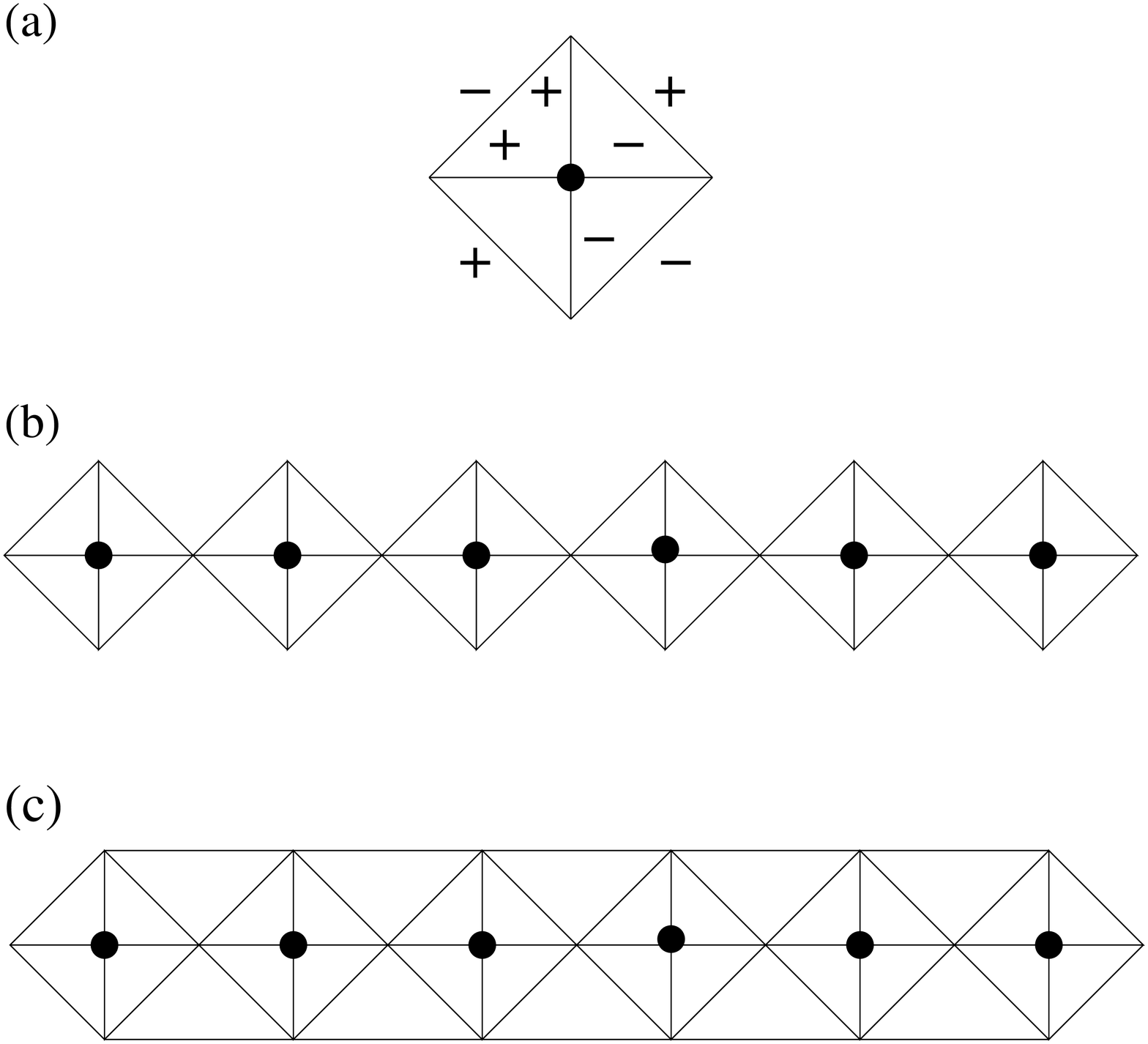}
\end{center}
\caption{ 
(a) Single unit cell of the $CuO_3$ system studied, the copper site is marked
by a filled circle and all vertices correspond to oxygen sites. 
The phase conventions  of the transfer
integrals, corresponding to the symmetry of the $d$- and $p$-orbitals 
are indicated  (b) Model 1 : 6 unit cells of the 
extended $CuO_3$ system with 
$t_{\parallel} = 0$ 
(c) Model 2 : similar to (b) with
$t_{\parallel} \ne 0$. 
  }
\end{figure}

In this paper we study currents in a 3-band extended Hubbard 
Hamiltonian on a
$CuO_3$-type system (Fig. 2), which is believed to model the essential
aspects of the 2D system. It is possible to envisage several quasi-1D
sections of the 2D $CuO_2$ plane. It is necessary to choose geometries that
are capable of supporting circulating currents.  The plaquette that
forms a single unit cell of  the $CuO_2$ plane is shown in Fig. 2a.
The system shown in Fig. 2b, henceforth to be called Model 1, consists 
of such plaquettes connected at a single point ($t_{\parallel}$ = 0).
A strip of the 
$CuO_2$ plane is shown in Fig. 2c. In this system, henceforth known as 
Model 2,  links parametrized by $t_{\parallel}$ have been added 
between adjacent plaquettes to model the  2D nature of the
system. This is expected to better correspond to scenario in the 
$CuO_2$ planes where 
the plaquettes are all coupled due to hopping along both axes. 

The essential difference between models 1 and 2 is that since the 
plaquettes of Model 1 are connected at just a single point, it is
possible to globally reverse the signs of the currents in each unit-cell
completely independently of all the other unit-cells, without any energy
cost. Therefore, while each unit-cell is a-priori capable of sustaining 
currents, long range order is not possible in Model 1.
Thus, it would be necessary to include links between these plaquettes
(present in the physical system), a scenario which corresponds to Model 2.
In this paper, we have examined  both these models in detail in order to study the
effects of interactions locally in a single unit cell as well as the 
long-range effects brought about by coupling plaquettes as in Model 2.

For weak interactions, ladder-type systems can be treated using
a combination of bosonization and perturbative renormalization group
techniques, as in the case of the 2-leg ladder (with disorder)\cite{orignac}
or  N-leg Hubbard ladder\cite{balents}. In fact, abelian bosonization
methods have been used to study staggered currents in the 2-leg 
ladder\cite{marston1}.  These powerful analytical tools have proved
useful in extracting a great deal of information. 
However, these methods cannot fully take into account the 
exact values of the
parameters as well as the precise geometry of the system, both
of which are crucial to the copper-oxides as discussed above. 
While our system appears to be
a quasi-1D system, it has a highly non-trivial geometry and 
up to $4^{th}$ neighbours are connected. 
Given the large number of parameters in the 3-band Hubbard
model and the importance of the delicate balance between them,
we have chosen to use the highly accurate DMRG method which gives
information even in the strong-coupling regime 
inaccessible to perturbative approaches.

We have used  a 3-band extended Hubbard model on the systems in
Fig. 2, described 
in greater detail below.
The 3-band Hubbard model for clusters of the $CuO_2$  system
(up to $6 \times 6$ unit-cells) has 
been studied by the constrained path quantum Monte Carlo method (CPMC)
\cite{gubernatis1,gubernatis2}.
However, the focus of 
the CPMC studies
was pair-binding of holes as well as the pairing and other
correlation functions, but not currents.
Thus, a study of flux-type phases or currents in this model does not exist, to
our knowledge.

\section{Model and Method}

The 3-band Hamiltonian for our $CuO_3$ system in the hole picture is given by:
\begin{eqnarray}
& & {} H  =  H_0 + H_1 + H_2  \\ 
& & {} H_0   = \Big[ \sum\limits_{i,\sigma} -t_{pd} d^{\dagger}_{i\sigma} 
(p_{x,i+a/2,\sigma}-p_{x,i-a/2,\sigma}  -
p_{y,i+a/2,\sigma}  \nonumber \\
& & {} +p_{y,i-a/2,\sigma})  -
t_{pp}p^{\dagger}_{x,i+a/2,\sigma}
(-p_{y,i+a/2,\sigma}+p_{y,i-a/2,\sigma})-  \nonumber \\ 
& & {} t_{pp}p^{\dagger}_{x,i-a/2,\sigma}
(p_{y,i+a/2,\sigma}-p_{y,i-a/2,\sigma})  \nonumber \\ 
& & {} +\sum\limits_{\langle ij \rangle \sigma} 
t_{\parallel} (p^{\dagger}_{y,i+a/2,\sigma}
p_{y,j+a/2,\sigma}  + 
p^{\dagger}_{y,i-a/2,\sigma} p_{y,j-a/2,\sigma}) \Big] \nonumber \\ 
& & {} + h.c. +
\sum\limits_{i\mu\sigma} (\epsilon_p n^i_{p_{x(y),\mu}\sigma}+
\epsilon_d n^i_{d\sigma}) \nonumber \\
& & {} H_1 =  
 U_d\sum\limits_{i\sigma}n^i_{d\sigma}n^i_{d{\bar\sigma}}+
   U_p\sum\limits_{i\sigma\mu}n^i_{p_{x(y),\mu}\sigma}n^i_{p_{x(y),\mu}{\bar\sigma}}  \nonumber \\
& & {}  H_2 =  V_{pd} \sum\limits_{i \mu} 
    n^i_{d} (n^i_{p_{x,\mu}}  + n^i_{p_{y,\mu}}) + 
  V_{pp} \sum\limits_{i \mu}  n^i_{p_{x,\mu}}n^i_{p_{y,\mu}} \nonumber \\
& & {} + 
 V_{\parallel} \sum\limits_{\langle ij \rangle \mu}  n^i_{p_{y,\mu}}
n^j_{p_{y,\mu}}
 \nonumber
\end{eqnarray}
where $d^{\dagger}_{i\sigma}$ creates a hole of spin $\sigma$
on the copper $d$-orbital at unit-cell $i$, $p_{x(y),i\pm a/2,\sigma}$
creates a hole of spin $\sigma$ at the oxygen $p_x$ ($p_y$) orbital at 
sites $\mu = i \pm a/2$  in the $x$ and $y$ directions of unit-cell $i$. 
Here $\langle ij \rangle  $ 
corresponds  to  a sum over nearest-neighbour unit cells $i$ and $j$
and the summation over $ \mu$ corresponds to $i \pm a/2$.
Operator $n^i_{d\sigma}$ is the number operator for the copper atom
of unit cell $i$ and 
$ n^i_{p_{x(y),\mu}\sigma}$ are the number operators for the 
$p_x$ ($p_y$) orbitals at sites $\mu = i \pm a/2$ of unit cell $i$.
The parameters  $\epsilon_d$ and $\epsilon_p$ are the site-energies
of the copper and oxygen sites ($\Delta = \epsilon_d - \epsilon_p$).
The copper-oxygen hopping is parametrized by  $t_{pd}$ and
 the oxygen-oxygen hopping by $t_{pp}$ while
$t_{\parallel}$ is the hopping between the $p_y$ orbitals
of neighbouring unit cells.
$U_d$ and $U_p$ are the on-site Hubbard repulsions at the 
copper  and oxygen sites, respectively.
The nearest-neighbour copper-oxygen repulsion 
is parametrized by $V_{pd}$, the oxygen-oxygen repulsions within a
unit cell by $V_{pp}$ and the repulsions between $p_y$ orbitals
in neighbouring unit cells by $V_{\parallel}$. 
Further, the phases of the hopping integrals $t$, corresponding to
the hybridization of $d$- and $p$-orbitals are represented 
in Fig. 2.

We study this model by the DMRG method. 
The complicated geometry (up to 4th neighbours are 
connected) of this system required the implementation of a
 7-block scheme (Fig. 3). This choice of RG scheme 
ensures that the relatively complicated geometry,
believed to be crucial 
in the cuprates, is preserved at each step of the calculation.
In Fig. 3a, we see the breakup of the two unit-cell system
which is the starting point of our calculation.
The system size increases by one unit cell (4 sites) at each iteration.
The blocks marked '1' and '7' in the figure are renormalized at each 
iteration and five new blocks (unrenormalized) of one site each 
are added to the middle
of the system. The dimension of the Hilbert space of each of the added sites is 4,
corresponding to empty, one hole (up and down spins) and 2 holes.
In Fig. 3b, we represent the blocks and superblocks of
the three unit-cell system. The following iterations can be obtained 
by replacing the superblocks '1' and '7' by the appropriate 
system.
The DMRG method is  a highly efficient way to study 1- and quasi 1-dimensional
strongly correlated fermionic systems. We are also not limited 
to a choice of fillings that correspond to closed-shells, a 
habitual constraint in quantum Monte Carlo simulations.

\begin{figure}
\begin{center}
\epsfxsize=0.8\columnwidth
\epsfbox{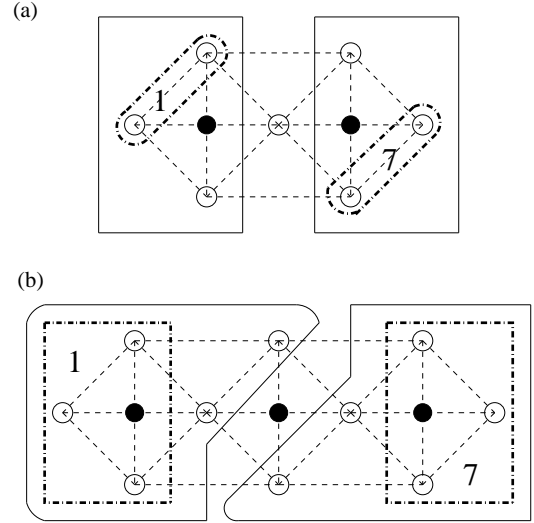}
\end{center}
\caption{The breakup of the extended system into blocks and superblocks used
in our 7-block  DMRG scheme for (a) even iterations  and (b) odd iterations. Copper
atoms are marked by filled circles and oxygen atoms by open circles.
The physical bonds are marked in dashed lines. The superblocks of each
iteration are marked in dot-dashed lines and also carry their 
number (1 and 7). The five central blocks consist of one site each.
 Superblock sizes increase 
by two sites in each iteration and the superblock of the 
following iteration is marked by solid lines.  }
\end{figure}

The parameters varied in our study are mainly the interaction strengths
($V$'s), $\Delta$ and the filling.
We have studied this model with
$ t_{pp}$ =  0.6,  $U_d$ = 10.0 and $U_p$ = 5.0 in units 
where $t_{pd} = 1.0$. Further, we consider two different geometries 
corresponding to values of $t_{\parallel} = 0$ and 
 $t_{\parallel} = 0.6$. The former corresponds to the
 case of plaquettes that  are coupled only at  a point (Model 1)
 and the
 latter to the case of fully connected plaquettes (Model 2).
 The choice of parameters for 3-band models of the copper-oxides
 has been extensively discussed in the literature \cite{parameters}.
Our choice corresponds to quite a 
realistic range of parameters for the copper oxides \cite
{gubernatis1,gubernatis2,parameters}.
In fact, one of the inherent difficulties associated with any
study of the copper-oxides is the
large number of parameters, giving rise to a
huge phase space. We have tried to vary the 
different $V$'s in order to locate the interesting 
part of the phase diagram.  One of the difficulties encountered in 
 our calculations has been the inaccessibility of certain
ranges of $V$ 
due to convergence problems, as discussed in the end of Section IIIB.

We have studied up to 25 $CuO_3$ unit cells, corresponding to
101 sites, at different dopings. 
Half-filling  corresponds to one hole per copper site
in the hole picture, where the system is known to be an antiferromagnetic
insulator for a wide range of parameter values. We have studied 
systems with different numbers of holes, corresponding to 
doping of the physical system.
We add two holes at every other iteration. Thus, the filling 
is not the same at every step of the calculation. The cases
we discuss are systems with 20 - 25 unit cells. 
 We have studied systems of $N$ unit cells with 
 $n_h$ holes given by, $n_h = N + 3 + mod(N+1 , 2)$.
The filling $x$ is given by $ n_h / N - 1$. Thus, $x$ = 0
corresponds to the half-filled case.
We have studied fillings $x$ = 0.12, 0.13 and 0.19 
 ($N=25, 23, 21$) for odd numbers of unit cells and
 $x$ =  0.17, 0.18 and 0.25 ($N = 24, 22, 20$) for 
 even numbers of unit cells.  Thus, we are able to study a 
 range of doping values. In addition, we have studied 
the case of $n_h = N + 1 + mod(N+1 , 2)$, for the model
with $t_{\parallel}$ = 0.

We compared the ground state energies for several iterations
with the energies obtained from exact diagonalization calculations to
verify our RG scheme.
We varied the number of states $m$ on the superblocks (
$m$ = $125,150$ and $200$).
We have kept up to 200 states on the superblocks.
This correspond to a Fock space dimension $H_F = m^2 \times 4^5$ =
40,960,000.
This provides 
sufficient accuracy, estimated in several ways. The discarded
weight of the density matrix states is of order  $10^{-6}$ with
200 states. In certain cases, we have also calculated the 
energies and properties of the non-interacting system (free-particle
model with all $U$'s and $V$'s set to zero), where exact results
are available. This provides a strong check on our data, since 
convergence is known to be most difficult in this case for the
DMRG algorithm.  Another property that we have used to check the
quality of our data is the magnetic moment $|\langle S^z_i \rangle | $,
at site $i$,
which ought to be zero in principle. Thus, only data which correspond to
relatively small magnitudes of $|\langle S^z_i \rangle | $ have been
used. 

\section{Results and Discussion}

The  physical property we have studied is   the
current-current (CC) correlation function for links $k$ and $l$.
The current along a bond connecting  sites $\mu$ and $\nu$ is defined as
$J_{\mu\nu} = it_{\mu\nu} \sum\limits_{\sigma} (c^{\dagger}_{\mu\sigma}
c_{\nu\sigma} - c^{\dagger}_{\nu\sigma}c_{\mu\sigma})$. 
The mean-field analysis of the 3-band Hubbard 
model indicated that the nearest-neighbour
repulsion $V$ is essential to stabilize the circulating-current
phase\cite{varma}. 
The circulating-current phase shown in Fig. 1 would then
consist of currents represented  by  the
arrows in the figure. Since we work in the real representation
of the Hamiltonian, we use the current-current correlation
functions as a measure of this order.

We compared our data for the non-interacting  model  with
the exact data
available in this case to assure ourselves of 
convergence.
The current for bond $l$ in unit-cell $i$
is denoted by $J^{i}(l)$, according to the conventions for 
bond numbers shown in
Fig. 4. We refer to the vertical copper-oxygen bonds as rungs and
the horizontal copper-oxygen bonds as legs in analogy to the
terminology used for ladders.

\begin{figure}
\begin{center}
\epsfxsize=\columnwidth
\epsfbox{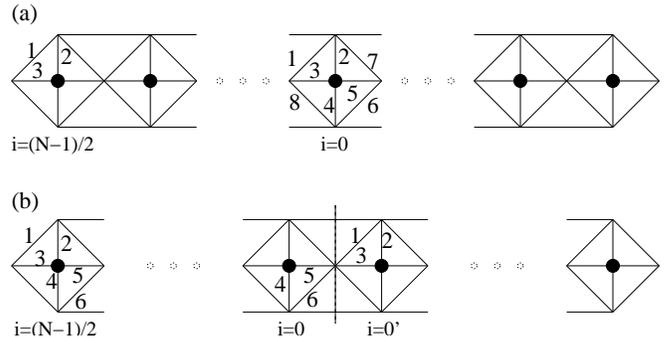}
\end{center}
\caption{Conventions for current-current correlation functions studied for 
systems with (a) odd  and (b) even numbers of unit cells. Unit-cell
numbers $i$ are marked beneath and the numbers in the figures
correspond to the bond numbers. }
\end{figure}

We have calculated the equal-time, zero-frequency current-current (CC)
correlation function, which is the ground state expectation value 
$\langle J^{0}(k) J^{i}(l) \rangle$. These correlations are between
currents in the central 
unit cell, denoted by $J^{0}(k)$ and  currents on the other unit-cells 
on one half of the system $J^{i}(l)$ ($i = 1, \ldots (N-1)/2$), for 
systems with odd numbers of unit-cells. 
Slightly different  correlations are
calculated at even and odd iterations due to the DMRG scheme used
and these are pictured in Fig. 4. 
For systems with even numbers of unit-cells,
we introduce the notation $\langle J^{0}(k) J^{i}(l) \rangle$, 
and $\langle J^{0^{\prime}}(k) J^{i}(l) \rangle$
with $i = 1, \ldots (N-1)/2$, as shown in Fig. 4.
However, we continue to refer to both cases as $\langle J^{0}(k) J^{i}(l)
\rangle$ in the text, with the different notations
for even and odd unit-cells being implicit.
By the bilateral symmetry about the $O-Cu-O$ axis,
it is not necessary to take into account all the possible cases.
We therefore restrict ourselves to the case of $\langle J^{0}(k) J^{i}(l)
\rangle$, $k$,$l$ = 1, \ldots 3.
We analyse cases of 
systems with 20-25 unit-cells, since these are sufficiently large 
to study the decay of the current-current correlations.
In particular, we analyze the
behaviour of the different types of diagonal ($\langle J^{0}(k) J^{i}(k)
\rangle$ and off-diagonal  $\langle J^{0}(k) J^{i}(l)\rangle$, 
($k \ne l$)
current-current correlations and try to classify them. 

\subsection{Analysis of Model 1}

We first present our results for the CC correlations
on the $CuO_3$ system with Model 1 ($t_{\parallel} = 0$) which consists 
of plaquettes coupled at a single point. We have performed calculations
with  values of $t_{pd}$ and $t_{pp}$ equal to 1.0 and 0.6. We have varied
$\Delta$ and $V_{pd}$, while $V_{pp}$ was set to 0.
We have further considered two sets of fillings corresponding to 
$n_h = N + 3 + mod(N+1 , 2)$ and $n_h = N + 1 + mod(N+1 , 2)$.
The second series then gives us  fillings $x$ = 0.04, 0.043 and 0.048
($N = 25, 23, 21$) and $x$ = 0.083, 0.09 and 0.1 ($N = 24, 22, 20$).
Thus, the second series $n_h = N + 1 + mod(N+1 , 2)$ has the
advantage of providing fillings that are practically constant
for different system sizes. This helps to verify that our conclusions
hold over different system sizes.

\begin{figure}
\begin{center}
\epsfxsize=0.8\columnwidth
\epsfbox{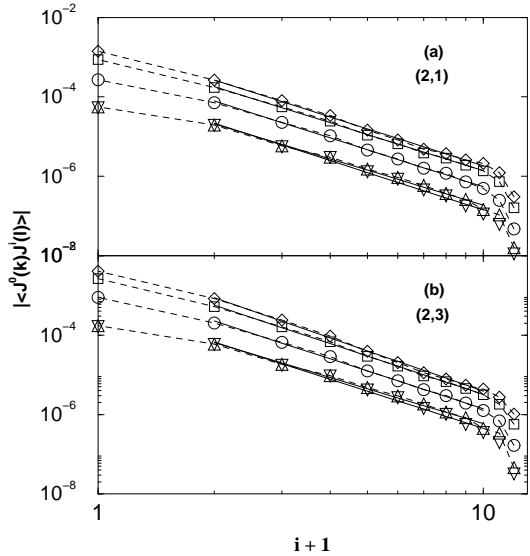}
\end{center}
\caption{ Current-current correlations $| \langle J^0 (k) J^i (l) \rangle |$
for a 25 unit-cell system with 28 holes, for Model 1, 
in a log-log representation.
All systems have the parameters
$t_{pp} = 0.6$, $t_{\parallel} = 0$, $\Delta=1$ 
in units with
$t_{pd}$ = 1.  
 Circles represent the case with the above parameters and
all $V$'s = 0, squares are for $V_{pd} = 1$ and
diamonds are for $V_{pd}  = 2$, all with $U_d = 10$ and $U_p = 5$,
obtained from DMRG calculations with $m = 200$ states. 
Triangles represent the exact 
values for the correlation for the non-interacting model 
($U_p = U_d = V_{pd} = V_{pp} = 0$), and inverted triangles  
the values obtained for the same parameters
from $m = 200$ state DMRG calculations, presented for comparison.
Dashed lines are a guide to the eye while full
lines correspond to the power-law fits of the data.
(a) represents the $| \langle J^0 (2) J^i (1) \rangle |$ 
correlation with $\alpha_{NI}^* = -2.94$, $\alpha_{NI} = -3.12$,
$\alpha_0 = -3.09$, $\alpha_1 = -3.03$, $\alpha_2 = -3.09$
and (b) the $| \langle J^0 (2) J^i (3) \rangle |$ correlation
with 
 $\alpha_{NI}^* = -2.92$, $\alpha_{NI} = -3.11$,
$\alpha_0 = -3.16$, $\alpha_1 = -3.21$, $\alpha_2 = -3.37$.
See text for definitions of the $\alpha$s.
  }
\end{figure}

We have also carried out numerical fits on our data. For our results on
Model 1, we found our data well-fitted by power-laws. We define
$\alpha_{NI}$ as the slope corresponding to the non-interacting case 
obtained from DMRG and  $\alpha_{NI}^*$ as the slope of the exact data.
Likewise $\alpha_{k}, k = 0\ldots2$ correspond to the cases with
$V_{pd} = k$. These slopes give us the exponents of the
power-law.

We have studied the effect of increasing the  copper-oxygen repulsion
$V_{pd}$ on the current-current correlations. In Fig. 4a, we present
results for the dependence of the CC correlation function
$| \langle J^0 (2) J^i (1) \rangle |$ with distance $i$ for a 25 unit-cell
system with 28 holes. From Fig. 4, it can be seen that this is the
correlation of a rung of the central unit-cell with an oxygen-oxygen
bond ($l$ = 1) of the unit-cells away from the centre.  We first 
consider this correlation for the non-interacting case (all $U$'s and 
$V$'s set to 0), where exact results are available. We have compared the
results obtained for $| \langle J^0 (2) J^i (1) \rangle |$ from
a DMRG calculation with $m = 200$ states with the exact results.
The convergence of the data is excellent, as seen in Fig. 5a.
The exponent of the DMRG data $\alpha_{NI} = -3.12$, compares 
favorably with the exponent obtained for the exact data, 
$\alpha_{NI}^* = -2.94$.
This is a strong check on our results, since the convergence 
of the DMRG method is known to be harder for the free-particle 
model than for the equivalent system with strong correlations.

We now note from Fig. 5a that the  curve which represents
current-current correlations for the
purely local Hubbard model with $U_d = 10, U_p = 5$ and $V$'s = 0,
lies above the non-interacting
curve. We then turn on nearest-neighbour interactions.
 Increasing the copper-oxygen repulsion $V_{pd} = 1$ 
has the effect of increasing 
the magnitudes of the current-current correlations. Further increasing
$V_{pd}$ to 2 increases the correlations, but by a smaller amount.
The decay of the correlations follows power-law behaviour, 
as seen from the log-log plots.  
Thus, we observe that
repulsive interactions are capable of increasing the
magnitude of the currents. It can be seen from the curves 
that the decay at large-distances is hardly affected by the
$V$s, the curves are nearly parallel. This is borne out by the
slopes ($-3.09, -3.03$ and $-3.09$), which are practically unchanged.
 This is because the plaquettes are coupled at a single point,
which prevents the formation of a state with long-range-order.

In Fig. 5b, we present our results for the correlation
$| \langle J^0 (2) J^i (3) \rangle |$. This is a correlation between
a copper-oxygen rung of the central unit-cell, with a 
copper-oxygen leg of the side unit-cells. Increasing the copper-oxygen
repulsion enhances the magnitude of the correlations compared to 
the Hubbard model, while
their decay is power-law, again with similar exponents,
as is seen from the curves which are nearly parallel and from the 
power-law fits.

In Fig. 6a we present the behaviour of the correlation $| \langle J^0 (2) J^i
(2) \rangle |$, which is the copper-oxygen rung-rung correlation
function. The data presented are for the case of 25 unit-cells with
26 holes, since the absolute magnitude of the correlations is
much larger than in the case of 28 holes. The qualitative behaviour remains
the same. We observe that the absolute values of the $V_{pd} = 2$
data are enhanced by almost an order of magnitude compared to the
$V_{pd} = 0$ curve. We have fitted our data with two series of power-laws
for even and odd sites in order to take into account the alternating behavior
of the curves. The slight differences between $\alpha_0$ and 
$\alpha_1$ can be attributed to the rather large alternance 
of the $V_{pd}$ = 0 data. This alternance is reminescent of the 
generic behavior of correlation functions in quasi-1D systems,
due to the harmonics of $k_F$. We notice that the magnitude of the
oscillations decreases when $V$ is increased. In quasi-1D systems,
similar behavior of spin-spin and charge-charge correlations is
interpreted as an increase in the value of the charge-stiffness
($K_{\rho}$)\cite{haldane}.

\begin{figure}
\begin{center}
\epsfxsize=0.8\columnwidth
\epsfbox{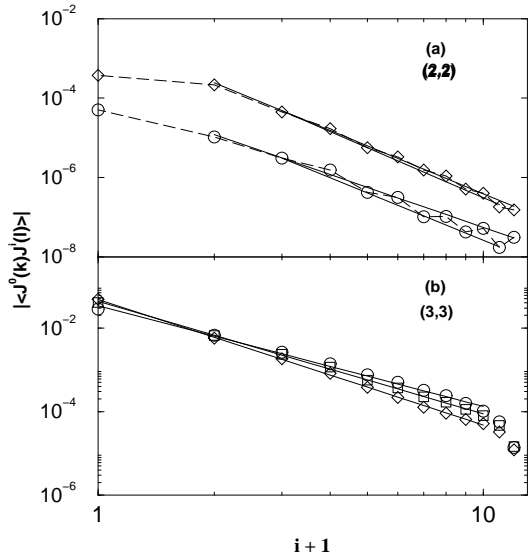}
\end{center}
\caption{  Current-current correlations for a 25 unit-cell system for
Model 1.  Note the log-log presentation.
All systems have the parameters
($t_{pp} = 0.6$, $t_{\parallel} = 0$, $\Delta=1$, $U_d = 10$, $U_p = 5$).
 Circles represent the case with the above parameters and
all $V$'s = 0, squares are for $V_{pd} = 1$ and
diamonds are for $V_{pd}  = 2$.
Dashed lines are a guide to the eye, while  full
lines represent the power-law fits.
(a) $| \langle J^0 (2) J^i (2) \rangle |$
for a 25 unit-cell system with 26 holes with
($\alpha_0 = -3.34$ (even) and $-3.94$ (odd));
($\alpha_1 = -4.00$ (even) and $-4.18$ (odd)).
and (b) $| \langle J^0 (3) J^i (3) \rangle |$ 
for a 25 unit-cell system with 28 holes with
($\alpha_0 = -2.40$, $\alpha_1 = -2.69$, $\alpha_2= -3.00$). 
  }
\end{figure}

The behaviour of the correlations $| \langle J^0 (3) J^i (3) \rangle |$
can be seen in Fig. 6b. This correlation is a copper-oxygen leg-leg
correlation function, as seen from Fig. 4.  Here, the behaviour 
observed is different from the previous cases studied, since the
correlation is enhanced very slightly
by increasing repulsions for first neighbours. 
However, the situation is reversed  at very short distances
and increasing repulsions causes
the correlations to decay even faster, as can be noted from the
crossing of the curves in Fig. 6b. Their decay is power-law 
with the exponents decreasing, as can be seen from the figure.

We have also analysed the behaviour of correlation functions
of the type $| \langle J^0 (1) J^i (1) \rangle |$. These 
are not presented for reasons of space. These correlations
are oxygen-oxygen correlations and decay as power-laws, with the magnitudes
falling as $V_{pd}$ is increased. This however, could be due to 
the value of $V_{pp}$ = 0, chosen here.

We now briefly discuss the case 
of $x = 0.18$ (22 unit-cells with 26 holes), which 
fits into the above analysis with an interesting 
reversal. In this case, the rung-rung correlations
(indeed, correlations involving one rung) are 
initially decreased when $V_{pd}$ is turned on. Increasing
$V_{pd}$ to 2 brings about a reversal of this behaviour, with
the magnitude of the correlations increasing compared to the
$V_{pd}$ = 0 case.
In Fig. 7 we present one example of this behaviour
with the correlation $| \langle J^{0^{\prime}} (2) 
J^i (3) \rangle |$, which is a rung-leg correlation. 
The reversal of the trend is clearly
seen from the  $V_{pd}$ = 1 curve (lies below the $V_{pd}$ = 0 curve)
and the $V_{pd}$ = 2 data which is enhanced with
respect to the $V_{pd}$ = 0 curve. This indicates that there
could be a threshold value of nearest-neighbour interactions
$V$ for enhancing the magnitude of current-current
correlations. 

\begin{figure}
\begin{center}
\epsfxsize=0.8\columnwidth
\epsfbox{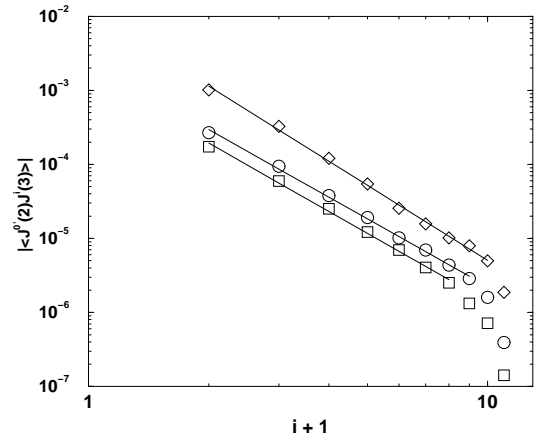}
\end{center}
\caption{ Current-current correlations 
$| \langle J^{0^{\prime}} (2) J^i (3) \rangle |$
for a 22 unit-cell system with 24 holes. Note the
log-log representation. All systems have the parameters
($t_{pp} = 0.6, t_{\parallel} = 0$, $\Delta=1$, $U_d = 10$, $U_p = 5$)
 Circles represent the case with the above parameters and
all $V$'s = 0, squares are for $V_{pd}  = 1$,
diamonds for $V_{pd} = 2$. Lines  represent power-law fits 
with exponents $\alpha_0 = -2.41$,  $\alpha_1 = -2.69$ and
 $\alpha_2 = -3.00$.
  }
\end{figure}

Thus, the above analyses show that the copper-oxygen rung-rung
correlation (or even correlations involving one rung) 
are enhanced in magnitude by repulsive interactions,
compared to the purely local Hubbard model, probably requiring a minimum
threshold value of $V$. 
The copper-oxygen leg-leg correlations show  an initial 
increase followed by a decay in magnitude and correlation 
length, as $V_{pd}$ is increased. The oxygen-oxygen 
correlation is diminished in magnitude and correlation 
length, as $V_{pd}$ is increased. We have analysed the 
cases of different fillings described above and have 
found similar behaviour in all these cases. 

\subsection{Analysis of Model 2}

We now turn to an analysis of Model 2 with $t_{\parallel} = 0.6$,
The behavior of this model is expected to be quite different from that
of the $t_{\parallel} = 0$ case. It is expected that the
long-distance behavior of the CC correlations is much more
sensitive to the nearest-neighbour repulsions $V$.
Here, we first study the purely local Hubbard model on our lattice
($U_d = 10, U_p = 5$ and $V's = 0$). We note that this corresponds to 
the system in Fig. 2c, with bonds connecting different plaquettes
($t_{\parallel} = 0.6$). However, the $V$ on this bond is initially
set to zero, to give us a model of connected plaquettes 
with purely local interactions. We the begin to turn on $V$s,
i.e we turn on copper-oxygen
and oxygen-oxygen nearest-neighbour interactions, along with
the interaction $V_{\parallel}$ ( $V_{pd} = V_{pp} = V_{\parallel} = 1.0$).
Thereafter, we further increase $V_{\parallel}$ to study the
effect of longer-range interactions on the current-current correlations.

We shall first consider the case of 25 unit cells at a filling of 0.12
in some detail before analysing our other data.
In Fig. 8a - 8c we present diagonal correlations of the type
$ |\langle J^0 (k) J^i (k) \rangle |$,
for $k$ =  1, \ldots 3, on a log-log scale. 
These correspond respectively to the oxygen-oxygen diagonal correlation, the
rung-rung correlation and the leg-leg correlation of bonds in the central 
unit cell with bonds in the side unit cells.
In Fig. 8a, we present the decay of the oxygen-oxygen diagonal correlation
($ |\langle J^0 (1) J^i (1) \rangle | $) with increasing
distance for increasing values of the nearest-neighbor repulsion $V$.
These are compared to the data 
with all $V$'s = 0, i.e. the purely local Hubbard model.
From the data in Fig. 8a, we note the effect of repulsions on
 the correlations
between currents on oxygen-oxygen bonds.
Turning on $V$'s, ($V_{pd} = V_{pp} = V_{\parallel} = 1$) causes the
correlations to decay in magnitude 
 compared to the Hubbard model. Further increasing
$V_{\parallel}$ to 3, causes the correlations to decay further. Data for
$V_{pd} = V_{pp} = 1, V_{\parallel} = 2$, fit smoothly into this 
pattern and have not been shown, for clarity. The 
correlations show exponential decay as can be seen from the
plots as well as the inset. We have fitted the Hubbard model 
curve and the $V_{\parallel} = 3$ data (scaled by a factor of 10)
by expressions of the form 
 $ y = A\,exp(\beta x) $, shown in the inset. The clear straight
 lines seen in the semilog plot are indicative of the quality
 of the fit. We obtain 
$A = -6.0065$ and $\beta = -0.6296$ for the Hubbard model ($V$s $= 0$) and
$A = -3.3280$ and $\beta = -0.7753$ for the $V_{\parallel} = 3$ data.

\begin{figure}
\begin{center}
\resizebox{7.5cm}{10cm}{\includegraphics{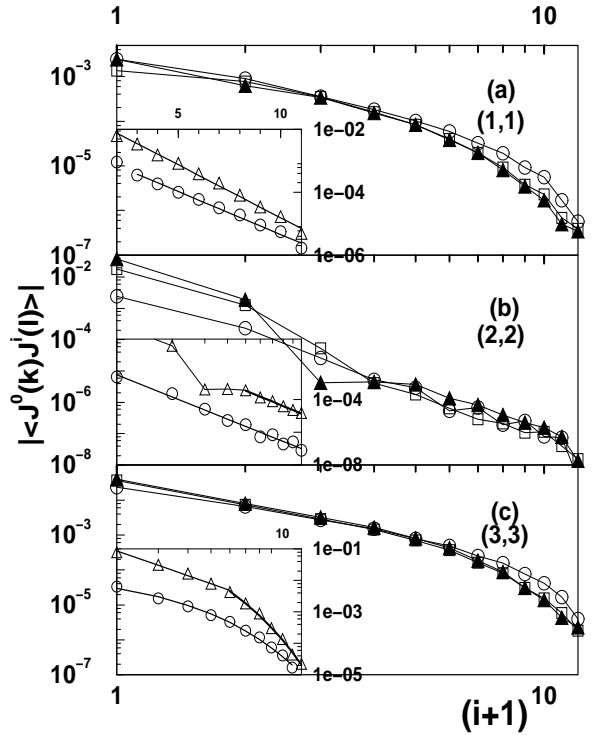}}
\end{center}
\caption{ 
Diagonal current current correlations 
for a 25 unit-cell system with 28 holes 
 ($|\langle J^0 (k) J^i (k)\rangle |$)
in a log-log representation. 
All systems have the parameters
($t_{pp} = t_{\parallel} = 0.6$, $\Delta=1$, $U_d = 10$, $U_p = 5$)
in units with
$t_{pd}$ = 1.  Circles represent the case with the above parameters and
all $V$'s = 0, squares are for $V_{pd} = V_{pp} = V_{\parallel} = 1$,
triangles for $V_{pd} = V_{pp} = 1, V_{\parallel} = 3$. 
Triangles are in some cases  filled, for clarity.
Lines in the main figures are meant to be a guide to the eye.
Lines in the insets correspond to fits which are discussed in greater
detail in the text.
From top to bottom, (a) diagonal oxygen-oxygen correlation function 
$ |\langle J^0 (1) J^i (1) \rangle| $. Inset (semilog representation)
contains $V_{\parallel}$ = 3 data multiplied by a factor 10, for visibility.
(b) rung-rung correlation  $ |\langle J^0 (2) J^i (2) \rangle| $. The inset 
(log-log plot)
contains $V_{\parallel}$ = 3 data multiplied by a factor 100. 
(c) leg-leg correlation $ |\langle J^0 (3) J^i (3) \rangle| $. 
The inset  (log-log plot)
contains $V_{\parallel}$ = 3 data multiplied by a factor 10. 
 }
\end{figure}

In  Fig. 8b we present 
cor\-re\-lations of type 
$ |\langle J^0 (2) J^i (2) \rangle | $, which correspond
to the copper-oxygen rung-rung correlation, 
for different values of interaction strengths.
Here, we notice a sizeable enhancement of the
current-current correlations (by an order of magnitude)
over distances of a few lattice 
spacings as interactions ($V$'s) are turned on, compared to the
Hubbard model.
 Thereafter, interactions cause 
 the correlations to fall, with respect to the $V$'s = 0 case.
However, further increasing $V_{\parallel}$
causes the correlations to rise i.e.   there is  a 
reversal of the trend.  
It is reassuring that the DMRG data in these cases 
are well converged, as seen from the small discarded weight and 
the  small
magnetic moments at the central sites, $ | S^z_i | \le 4 \times 10^{-4} $.
These data have been fitted by power-laws. The fits are good
as seen from the straight-lines in the log-log plots, both
in the main figure and in the inset (thick lines correspond to the
fits). The slope of the straight line
gives the exponent of the power-law and we obtain
 $\alpha_0  = -4.4070$ and $\alpha_3  = -4.6830$.
We also note that the cross-rung-rung correlation, of type 
$ |\langle J^0 (2) J^i (4) \rangle| $ are similar 
qualitatively and quantitatively to the rung-rung correlations
and are not presented here.

In  Fig. 8c we present 
correlations of type $ |\langle J^0 (3) J^i (3) \rangle | $, which correspond
to the copper-oxygen leg-leg correlation, 
for different values of interaction strengths.
Here, we notice a slight enhancement of the
current-current correlations over distances of a few lattice 
spacings as interactions ($V$'s) are turned on, compared to the
Hubbard model.
 Thereafter, interactions cause 
 the correlations to fall, with respect to the $V$'s = 0 case,
upon increasing $V$s.
The long-distance behaviour in this case is interesting. The 
$V = 0$ data (Hubbard model) are clearly exponential as seen
from the main figure and the inset 
($A = -3.9204$ and $\beta = -0.6254$). On the other
hand the $V_{\parallel} = 3$ data (triangles)  show much more
complex behaviour corresponding to  power-law behavior at short 
distances
(up to 6 sites) followed by an exponential tail at long distances.
The power-law fit is indicated in the inset by the thin line and
the exponential tail by the thick line. We 
obtain  the power-law exponent $\alpha_3 = -2.6299$ from the slope
of the power-law part of the fit and 
$A = -0.0143$ and $\beta  = -0.8947$ for the exponential part
(thick line).

We now turn to an analysis of the off-diagonal correlations.
Given the large number of correlations, we attempt to 
set out the general trends by treating one off-diagonal 
correlation of each type.
In Fig. 9a - Fig. 9c, we present off-diagonal CC correlations,
$ |\langle J^0 (1) J^i (2) \rangle |$,
$ |\langle J^0 (3) J^i (2) \rangle |$ and 
$ |\langle J^0 (3) J^i (1) \rangle |$.
The first of these corresponds to the correlation of an
oxygen-oxygen bond in the central unit-cell with
copper-oxygen rungs of the side unit cells. We refer 
to this henceforth as an "oxygen-rung" correlation.
As seen from the log-log plot of Fig. 9a, the correlation (1,2) 
exhibits power-law decay both for the Hubbard model and upon increasing 
$V$s. There is a substantial increase of this correlation 
at short distances with increasing $V$. This trend persists 
over long distances for the case of $V_{\parallel} = 3$ as seen 
from the figure. The power-law fits to these data are presented
in the inset, with the $V=3$ data scaled by a factor 10. 
We have fitted the Hubbard model data in
two series corresponding to the odd and even points to take care
of the alternance present in our data and obtain 
$\alpha_0$s of $-3.9408$ (odd) and $-3.3321$ (even). The
$V_{\parallel} = 3$ data were fitted by a power-law with exponent
$\alpha_3 = -4.0923$. We note that the alternance of the
$V_{\parallel} = 3$ is strongly diminished with respect to
the $V = 0$ data.

Similar behavior is observed in the leg-rung correlation function
 $ |\langle J^0 (3) J^i (2) \rangle |$, presented in Fig. 9b.
 The $V_{\parallel} = 3$ data are clearly enhanced with respect to the 
 Hubbard model data almost  over all the length of the curve. 
 We followed a procedure similar  to the previous case, fitting
 the $V = 0$ data by power-laws in two series
 to obtain $\alpha_0$s of $-3.0559$ (odd) and $-4.0292$ (even).
 The power-law fits to the $V_{\parallel} = 3$ data gave us an
 exponent $\alpha_3$ of $-4.1966$. Correlations quench the
 even-odd alternation of the data.

\begin{figure}
\begin{center}
\resizebox{7.5cm}{10cm}{\includegraphics{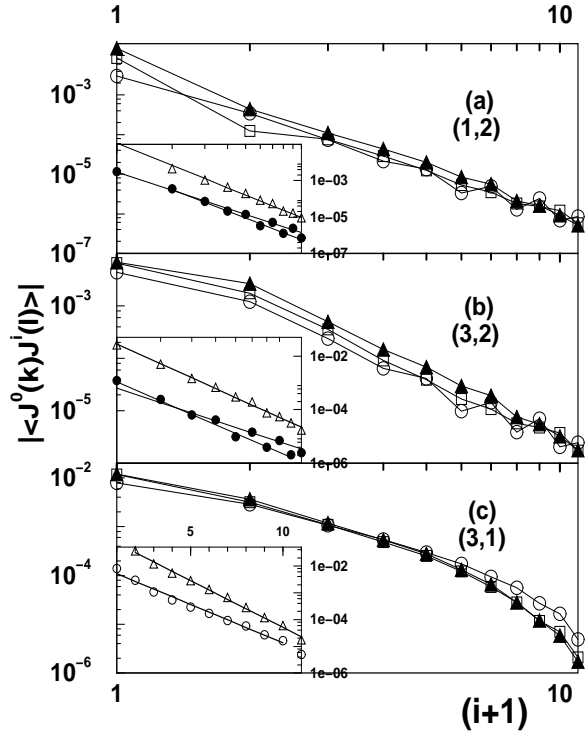}}
\end{center}
\caption{ 
Off-diagonal current current correlations 
$| \langle J^0 (k) J^i (l) \rangle |$ ($k \ne l$)
for a 25 unit-cell system with 28 holes in a log-log representation. 
All systems have the parameters
($t_{pp} = t_{\parallel} = 0.6$, $\Delta=1$, $U_d = 10$, $U_p = 5$)
in units with
$t_{pd}$ = 1.  Circles represent the case with the above parameters and
all $V$'s = 0, squares are for $V_{\parallel} = V_{pp} = V_{\parallel} = 1$,
triangles for $V_{pd} = V_{pp} = 1, V_{\parallel} = 3$. 
Symbols are in some cases  filled, for clarity.
Lines in the main figures are meant to be a guide to the eye.
Lines in the insets correspond to fits which are discussed in greater 
detail in the text.
From top to bottom, (a) "oxygen-rung" correlation
$ |\langle J^0 (1) J^i (2) \rangle| $. Inset (log-log)
contains $V_{\parallel}$ = 3 data multiplied by a factor 10, for visibility.
(b) leg-rung correlation  $ |\langle J^0 (3) J^i (2) \rangle| $. The inset 
(log-log plot)
contains $V_{\parallel}$ = 3 data multiplied by a factor 10. 
(c) "leg-oxygen" correlation $ |\langle J^0 (3) J^i (1) \rangle| $. 
The inset  (semilog plot)
contains $V_{\parallel}$ = 3 data multiplied by a factor 10.
  }
\end{figure}

In Fig. 9c, we present the correlation  
$ |\langle J^0 (3) J^i (1) \rangle |$, which corresponds to the 
correlation of the central oxygen-oxygen bond with the
copper-oxygen leg of the side unit cell. We refer to this
correlation henceforth as the "oxygen-leg" correlation. We 
observe from the figure and from the fits in the inset that
these correlations have exponential behavior for 
the Hubbard model as well as for the case with $V_{\parallel}=3$.
From the fits seen in the inset (presented in semilog form)
we obtain $A = -4.61476$ and $\beta = -0.6600$ for the 
$V = 0$ case and $A = -1.849094$ and $\beta = -0.8081$ for the
$V_{\parallel} = 3$ case.

\begin{figure}
\begin{center}
\epsfxsize=0.8\columnwidth
\resizebox{7.0cm}{9cm}{\includegraphics{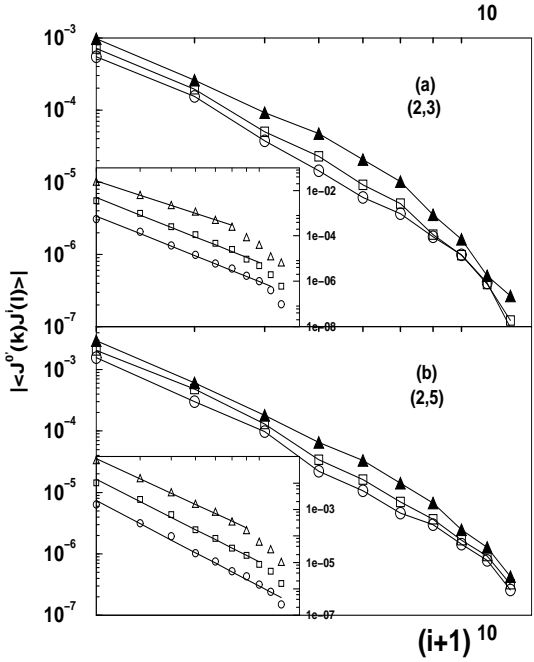}}
\end{center}
\caption{ Current-current correlations 
$| \langle J^{0^{\prime}} (2) J^i (k) \rangle |$
for a 22 unit-cell system with 26 holes. All systems have the parameters
$t_{pp} = t_{\parallel} = 0.6$, $\Delta=1$, $U_d = 10$, $U_p = 5$
in units with $t_{pd}$ = 1.  
Circles represent the case with the above parameters and
all $V$'s = 0, squares are for $V_{pd} = V_{pp} = V_{\parallel} = 1$
 and
triangles for $V_{pd} = V_{pp} = 1, V_{\parallel} = 3$. 
Triangles are in some cases  filled, for clarity.
Lines in the main figures are meant as a guide to the eye and lines
in the insets are the power-law fits to the data.
In both insets, $V_{\parallel} = 1$
data (squares) is multiplied by a factor of 5 and $V_{\parallel} = 3$ 
 data (triangles) by a 
factor of 25 for clarity. From top to bottom,
(a) rung-leg correlation, $| \langle J^{0^{\prime}} (2) J^i (3) \rangle |$.
From the fits, $\alpha_0 = -4.4157$, $\alpha_1 = - 4.3648$ and
$\alpha_2 = -3.5904$.
(b) another rung-leg correlation
$| \langle J^{0^{\prime}} (2) J^i (5) \rangle |$.
From the fits, $\alpha_0 = -4.9475$, $\alpha_1 = - 4.7718$ and
$\alpha_2 = -4.3446$.
}
\end{figure}

In addition to the data presented in
Figs.~8-9, for the 25 unit-cell system with filling $x = 0.12$,
we  discuss some results for the case of 22 unit-cells with
$x$ = 0.18.  
In Fig. 10, we present the evolution of  
two types of rung-leg correlations
that are similar: 
$ |\langle J^{0^{\prime}} (2) J^i (3) \rangle| $
and  $| \langle J^{0^{\prime}} (2) J^i (5) \rangle |$, henceforth
referred to as the $(2^{\prime},3)$ and $(2^{\prime},5)$
correlations respectively.
In Fig. 10a, we see that the $V_{\parallel} = 1$ data of the 
 $(2^{\prime},3)$ correlation  
 is already enhanced with respect to 
the Hubbard model data over intermediate distances. The
$V_{\parallel} = 3$ data is clearly enhanced with respect to the
Hubbard model data. We have  fitted these data by means
of power-laws, as seen from the log-log scale of the curve.
The results of the fits are seen from the inset. We see that $\alpha_3$,
the exponent of the $V_{\parallel} = 2$ curve is visibly reduced to
$-3.59$ compared to $\alpha_0 = -4.94$ for the Hubbard model
and the $V_{\parallel} = 3$ data falls off much more slowly than the
Hubbard model data.
As seen from the inset, the number of points we were able to
include in the fit decreases with increasing $V$, as 
compared to the Hubbard model.
 It is accepted that 
convergence of the DMRG algorithm is best in the presence
of strong, local interactions. Our Hubbard model case falls
into this category with $U_d = 10$ and $U_p = 5$. Increasing 
$V$ clearly renders convergence much more difficult, especially 
when $V_{\parallel} = 3$ which is most probably the reason
for this behavior.

In Fig. 10b, we present in a manner analogous to Fig. 10a, data 
for the $(2^{\prime},5)$
correlation functions. As in the previous case, we note that
the $V_{\parallel} = 3$ data is stabilized compared to the
$V = 0$ data and the $V_{\parallel} = 1$ data, with the exponents 
$\alpha_0 = -4.9475$, $\alpha_1 = -4.7718$ and $\alpha_3 = -4.3446$.
Thus, we note  that the decay of the   $(2^{\prime},5)$ correlation 
is slower for  $V_{\parallel} = 3$ than for the Hubbard model. We have
dropped 3 points from the $V_{\parallel} = 3$ curve and 1 point from 
the Hubbard model curve to obtain these results.

Thus, the analysis of the diagonal and off-diagonal
CC correlation functions
of Model 2 yields the following trends : the nearest-neighbor
repulsive interaction $V$ is clearly capable of increasing the
magnitude of the current-current correlations locally and
of strongly modifying the long-distance behavior. 
The correlations fall into different categories. 

The rung-rung correlation
function is enhanced by increasing $V$s and shows power-law decay.
In certain cases, the exponent of the power-law decreases upon 
increasing $V$ and the correlations fall off slowly as
compared to the Hubbard model.
In fact, this extends to any correlation that includes at least
one rung, diagonal or off-diagonal.

The leg-leg correlations show exponential decay in the Hubbard model.
However, increasing the values of $V$s and $V_{\parallel}$ in particular
changes this behaviour - we now have a power-law behavior at 
short distances followed by an exponential tail. Thus, we see
that $V$s are capable of locally enhancing the CC correlations and
by connecting the plaquettes, the long-distance behaviour is
greatly modified. 
We believe that the difference between the
rung-rung and leg-leg correlations is due to our choice of model
and that this difference should disappear in the 2D system.

The oxygen-oxygen correlation function shows exponential decay
with very slight enhancement of the absolute values for $V_{\parallel} = 3$. 
This is also true of the "oxygen-leg" type correlation functions.
This could however be due to our choice of parameter values. 
Clearly, it would be interesting to study other parameter
ranges for the Vs, especially for $V_{pd}$ and $V_{pp}$. This 
point is discussed further down in this section.

We have studied various cases involving other parameter values. Notably,
we tried to increase $V_{pd}$ and $V_{pp}$
($V_{pd}$ = 2.0, 3.0, $V_{pp}$ = 2.0, in units of $t_{pd}$),
but encountered convergence problems at the level of the Lanczos
diagonalization algorithm.  It is quite likely that  
increased degeneracy of the ground state
slows down convergence, as is known to occur in other cases 
such as the $J_1 - J_2$ spin chain, at parameter values close
to the critical point.
This  is most probably due to the increased delocalization of the 
ground state, brought about by larger $V$'s.
To overcome this problem,  it would be necessary to increase 
the number of states $m$ that we keep on
the superblocks. 
However,  at our present
value of $m$ = 200, we are already at the limit of our computational
resources. 
The charge densities in our
systems show that the charges tend to be concentrated
on the copper atoms, which possibly competes with the
stabilization of currents. Thus, increasing $V_{pd}$  and
$V_{pp}$ is likely to bring about a rehomogenization 
of charge and this could be favourable to the current-current correlations.

At this juncture, it is interesting to compare the models with
$t_{\parallel} = 0$ and 0.6. We have observed  enhancement in
the magnitudes of the currents in both models. Thus, repulsive 
nearest-neighbour interactions are clearly capable of increasing
the magnitudes of currents. The long-distance behaviour of 
correlations is different in the two models. 
Model 1  shows power-law decay of correlations and the interaction $V$ has
a very small effect on the exponents. 
In Model 2, any correlation involving a rung 
shows power-law behaviour.  We have very interesting indications
that repulsive nearest-neighbour interactions 
can enhance currents over distances comparable to the length scale 
of the system.  In certain cases, the exponent is
considerably reduced  and the current-current correlations
stabilized compared to the Hubbard model.

\section{Conclusions}
In conclusion, we have studied a 3-band extended Hubbard 
Hamiltonian on  a model  $CuO_3$ system, with  geometry
similar to that of a section of  the $CuO_2$ planes of the 
copper-oxides.  We have studied two different cases, with 
the parameter $t_{\parallel} = 0$  and 0.6,
which correspond respectively to Model 1 with plaquettes
coupled at a single point and Model 2, with plaquettes 
coupled all along the length of the system.
 Our calculations show that 
nearest-neighbour repulsive interactions are clearly
capable of enhancing current-current correlations
in magnitude and of qualitatively changing the long-distance behavior 
of the system.
Model 1  shows power-law decay of correlations 
and the interaction $V_{pd}$ has
a very small effect on the exponents.
Increasing the copper-oxygen repulsion $V_{pd}$ can
increase the magnitude of the  rung-rung current-current correlation
(or correlations involving a single rung)
by up to an order of magnitude.
In Model 2, 
the effect of introducing 
nearest-neighbour interactions $V$ in the system is to produce a
sizeable increase in the rung-rung correlations over  distances
of a few lattice spacings. 
$V_{\parallel}$ can enhance  correlations involving at least one
copper-oxygen rung over distances comparable to 
the length of the system. In certain cases, interactions bring about
 a sizeble decrease in the exponent, causing the correlations to
 decay slowly with respect to the Hubbard model.
We have indications that threshold values of the nearest-neighbour
repulsions are necessary to observe stabilization at long 
distances.  Tuning the 
copper-oxygen and the oxygen-oxygen repulsions  would
redistribute charge more uniformly in the system which is likely to
stabilize the current-current correlation functions in this
system. We believe that the enhancement and stabilization of
certain correlation functions seen in our model constitutes a
trend that will hold for the 2D system.

\noindent
{\bf Acknowledgements}: We thank the IDRIS, Orsay for time on the NEC SX-5
and the CICT, Toulouse for time on CALMIP.
We thank C. J. Calzado, J-P. Malrieu and C. M. Varma
for useful discussions.

\end{document}